\documentclass{aa}

\usepackage{graphicx}
\usepackage{pjournal}
\usepackage{mamd}
\usepackage{natbib}
\usepackage{amssymb}

\begin{document}

\title{Statistical properties of dust far-infrared emission}

 \author{M.-A. Miville-Desch\^enes\inst{1,2}\and G. Lagache\inst{1}\and
F. Boulanger\inst{1}\and J.-L. Puget\inst{1}}
 \institute{Institut d'Astrophysique Spatiale, Universit\'e Paris-Sud, 
B\^at. 121, 91405, Orsay, France \and Canadian Institute for Theoretical Astrophysics, 
University of Toronto, 60 St. George Street, Toronto, ON, M5S~3H8, Canada}

 \offprints{Marc-Antoine Miville-Desch\^enes}
 \mail{mamd@ias.u-psud.fr}
\date{\today}

\titlerunning{Statistical properties of dust FIR emission}
\authorrunning{Miville-Desch\^enes, M.-A. et al.}

\abstract
{
Far-infrared dust emission has a self-similar structure which reveals the complex dynamical processes
that shape the interstellar medium. The description of the statistical properties of this emission
gives important constraints on the physics of the interstellar medium but it is also a useful way
to estimate the contamination of diffuse interstellar emission in the cases where it is considered a nuisance.
}
{
The main goals of this analysis of the power spectrum and non-Gaussian properties of far-infrared
dust emission are 1) to estimate the power spectrum of interstellar matter density in three dimensions,
2) to review and extend previous estimates of the cirrus noise due to dust emission and 3) 
to produce simulated dust emission maps that reproduce the observed statistical properties.
}
{
To estimate the statistical properties of dust emission we analyzed the
power spectrum and wavelet decomposition of 100~\um IRIS data (an improved version of the IRAS data) 
over 55~\% of the sky. The simulation of realistic infrared emission maps is based on
modified Gaussian random fields.
}
{
The main results are the following. 1) The cirrus noise level as a function of brightness
has been previously overestimated. It is found to be proportional to $<I>$ instead of $<I>^{1.5}$,
where $<I>$ is the local average brightness at 100~\ump. 
This scaling is in accordance with the fact that the brightness fluctuation level
observed at a given angular scale on the sky is the sum of fluctuations of increasing 
amplitude with distance on the line of sight.
2) The spectral index of dust emission at scales between 5 arcmin and 12.5$^\circ$ is $<\gamma>=-2.9$
on average but shows significant variations over the sky. Bright regions have systematically 
steeper power spectra than diffuse regions.
3) The skewness and kurtosis of brightness fluctuations is high, indicative of strong
non-Gaussianity. Unlike the standard deviation of the fluctuations, the skewness
and kurtosis do not depend significantly on brightness, except in 
bright regions ($>10$~MJy~sr$^{-1}$) where they are systematically higher, probably due to contrasted
structures related to star formation activity.
4) Based on our characterization of the 100~\um power spectrum we provide a prescription of
the cirrus confusion noise as a function of wavelength and scale. 
5) Finally we present a method based on a modification of Gaussian random fields 
to produce simulations of dust maps which reproduce the power spectrum and non-Gaussian properties of
interstellar dust emission.
}
{}

\keywords{Methods: statistical -- ISM: structure -- Infrared: ISM: continuum -- dust}

\maketitle

\section{Introduction}

The interstellar medium emission shows fluctuations at all observable scales, 
revealing the self-similar nature of its density structure. 
The physical processes responsible for this self-similarity of the interstellar medium (ISM) structure
is yet to be fully identified. It could be related to turbulent motions
but also to chemical and thermal instabilities which trigger phase transitions
and play an important role in shaping the medium.

Faced with the challenge to understand and characterize the 
great structural complexity of interstellar emission, 
astronomers have used several statistical tools 
(power spectrum, correlation and structure functions,
wavelets, area-perimeter relation, principal component analysis,...) 
on several interstellar tracers : molecules 
\cite[]{falgarone1991,hobson1992,stutzki1998,bensch2001,brunt2003,padoan2003},
atomic hydrogen 
\cite[]{crovisier1983,green1993,stanimirovic1999,stanimirovic2001,dickey2001,elmegreen2001,miville-deschenes2003c},
extinction \cite[]{padoan2002,padoan2006} or
dust emission \cite[]{gautier1992,abergel1996a,jewell2001,ingalls2004}.

\begin{figure}
\includegraphics[width=\linewidth, draft=false]{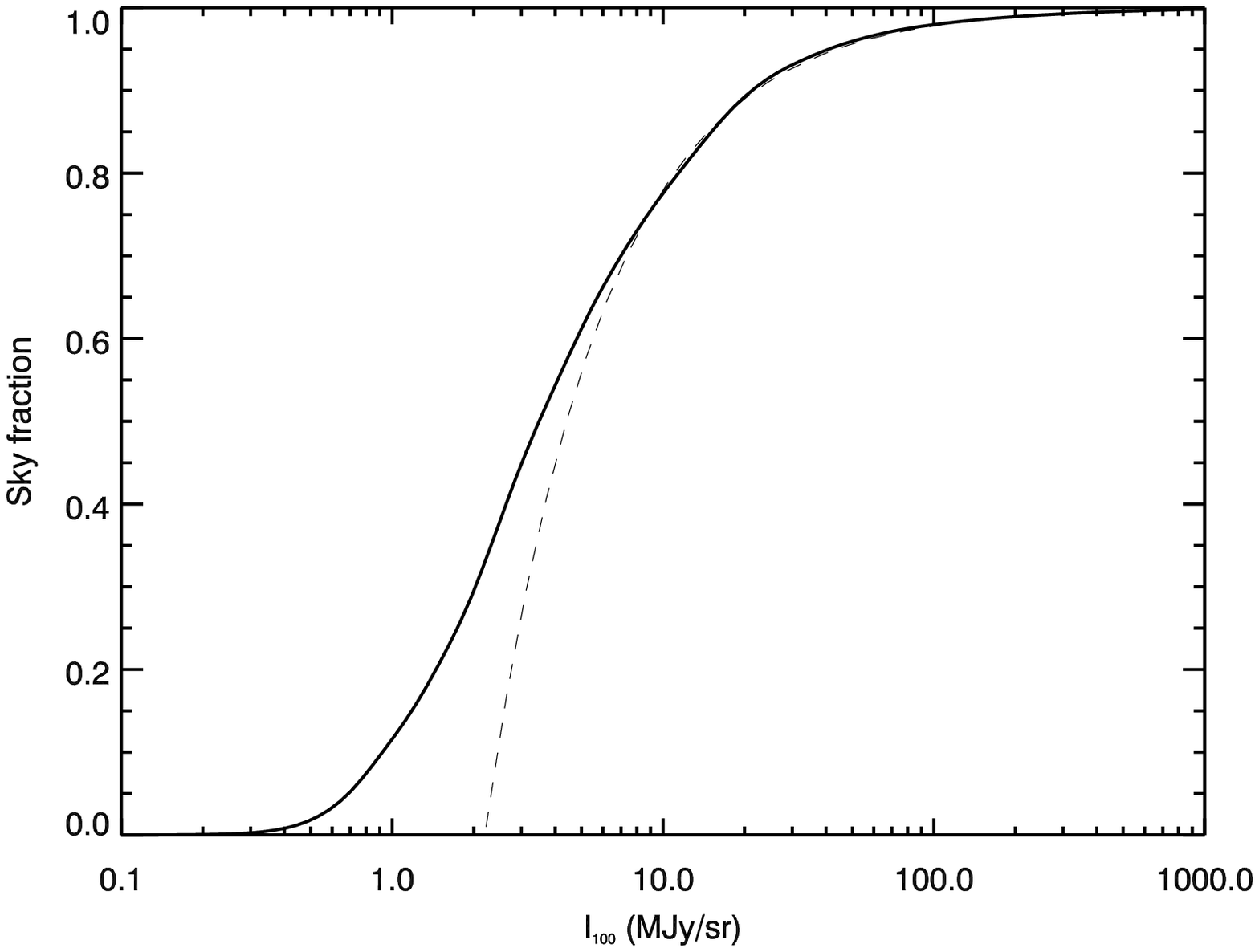}
\includegraphics[width=\linewidth, draft=false]{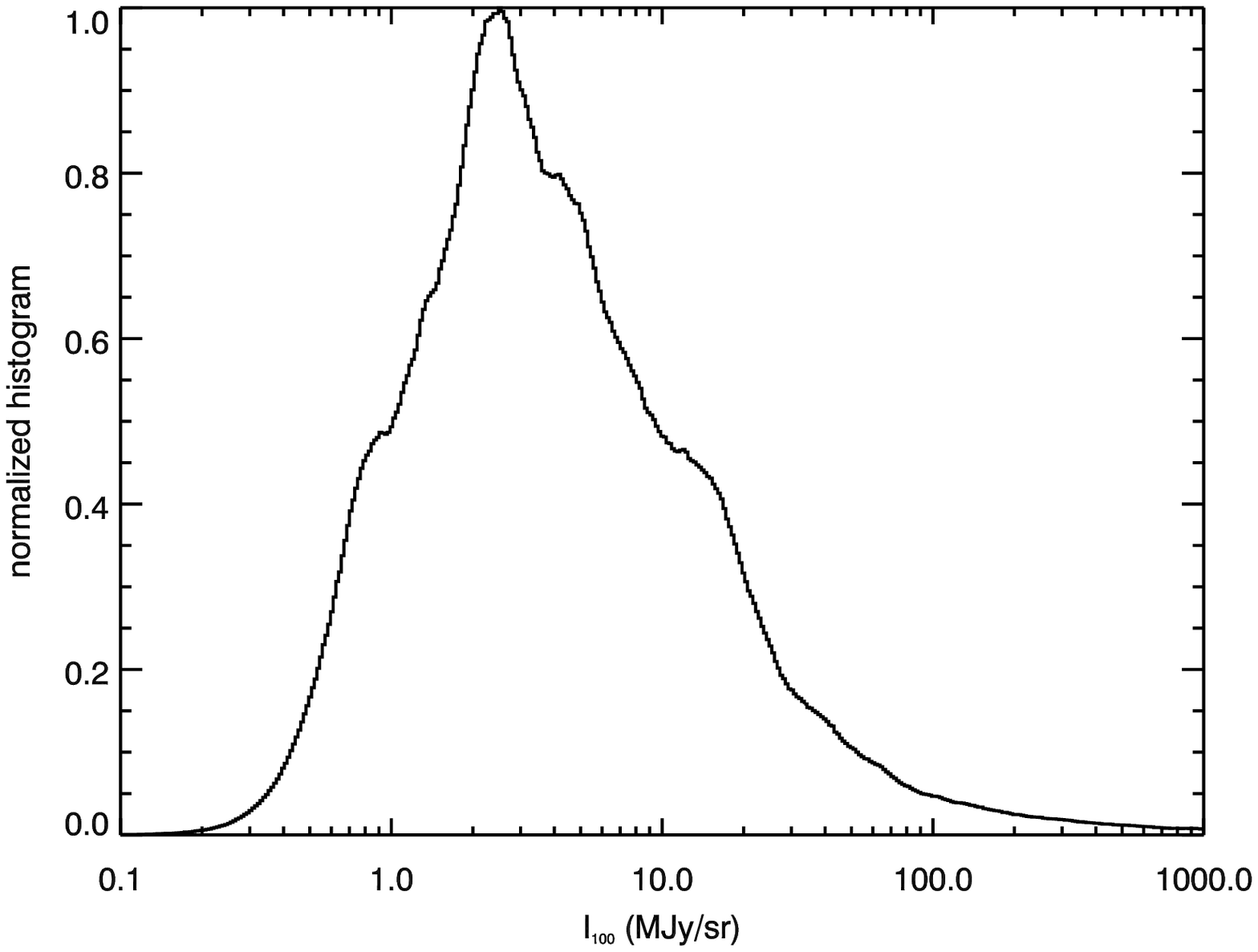}
\caption{\label{fig_cumulativeHisto} {\bf Top.} Cumulative histogram of 100~\um surface brightness
over 98\% of the sky (solid line): fraction of the sky with a brightness lower or equal to $I_{100}$.
The dashed line represent what would be expected from a cylindrical disk (cosecant law).
{\bf Bottom.} Histogram of $I_{100}$ brightness over the whole sky with bin scaled logarithmically.
For both curves we used the IRIS data projected on the Healpix grid (equal area pixels - http://healpix.jpl.nasa.gov)
with a pixel size of 1.7 arcmin (nside=2048). A background of 0.78 MJy~sr$^{-1}$ was removed
from the data to take into account the cosmic infrared background and any zodiacal light residual
\cite[]{hauser1998,lagache2000a}.}
\end{figure}

\begin{table}
\begin{center}
\begin{tabular}{cc}\hline
$<I_{100}>$ & 16.18 MJy sr$^{-1}$\\
median $I_{100}$ & 3.49 MJy sr$^{-1}$\\
most probable $I_{100}$ & 2.51 MJy sr$^{-1}$\\ \hline
$<0.1$ MJy sr$^{-1}$ & 0.006\%\\
$<0.2$ MJy sr$^{-1}$ & 0.06\%\\
$<0.5$ MJy sr$^{-1}$ & 1.8\%\\
$<1.0$ MJy sr$^{-1}$ & 11.6\%\\
$<2.0$ MJy sr$^{-1}$ & 29.5\%\\
$<5.0$ MJy sr$^{-1}$ & 61.1\%\\
$<10.0$ MJy sr$^{-1}$ & 77.6\%\\
$<20.0$ MJy sr$^{-1}$ & 89.2\%\\
$<50.0$ MJy sr$^{-1}$ & 95.6\%\\\hline
\end{tabular}
\caption{\label{table:stat} Average, median and most probable value of the 100~\um brightness
over 98~\% of the sky at an angular resolution of 4.3 arcminutes. The bottom portion of the table
gives the fraction of the sky with brightness lower than the value given in the left column.}
\end{center}
\end{table}

The analysis of the results of these tools is made difficult by projection, 
and instrumental effects but also by the fact that no observation is a
perfect tracer of the total gas column density. 
Several works have been dedicated to the understanding of these effects 
\cite[]{goldman2000,elmegreen2001,padoan2001,miville-deschenes2003b} 
and specifically on how one can retrieve
the three-dimensional statistical properties of interstellar matter 
from astronomical observations. 
The recent theoretical progresses made in that area open interesting perspectives on
the use of statistical tools to determine the three-dimensional structure
of the gas and to identify privileged scales at which physical processes
are important. 

The characterization of the statistical properties of
the interstellar emission is relevant for the understanding of
the physics of the ISM but is also of importance for studies where 
interstellar emission is a nuisance. This is the case of the study of pre-stellar
cores in molecular clouds but also of the analysis of the
Cosmic Microwave Background (CMB) and Cosmic Infrared Background (CIB)
radiations. In these cases the interstellar emission is considered
as a noise (the so-called ``cirrus noise'') for which one needs a 
detailed statistical description in order to remove it and account 
for it in the error budget. 
In that respect the interstellar medium is a rather complex noise source
due to its highly non-white and non-Gaussian brightness fluctuations.

The present paper is a study of the power spectrum of interstellar dust emission
at 100~\ump. The main goals are to use such analysis to put some constrains on the density structure of the interstellar medium
but also to propose a caveat for estimation and simulation of cirrus noise.
\cite{gautier1992} showed that the power spectrum of 
the IRAS 100~\um emission is characterized by a power law 
$P(k) = A \, k^\gamma$
where the exponent $\gamma\sim -3$ is independant of the brightness.
Moreover \cite{gautier1992} showed that the normalisation
factor of the power spectrum depends on the mean brightness $<I>$ of the 
region considered, with $A \propto <I>^3$.
This study has been conducted on a relatively small number of regions
and at a time where the instrumental response of IRAS was not
well known. In this paper we would like to revisit the work of
\cite{gautier1992} by extending it to the whole sky 
and by taking advantage of the recent reprocessing of the ISSA plates
by \cite{miville-deschenes2005a}.

In \S~\ref{data} we present the data used in this analysis.
The results of the power spectrum analysis and the implication on the ISM density structure
are presented in \S~\ref{power_spectrum_analysis}.
An estimate of the cirrus noise level in the FIR-submm is provided in \S~\ref{cirrus_noise}
and a method to produce realistic dust emission maps is given in \S~\ref{artificial_maps}.

\begin{figure}
\includegraphics[width=\linewidth, draft=false]{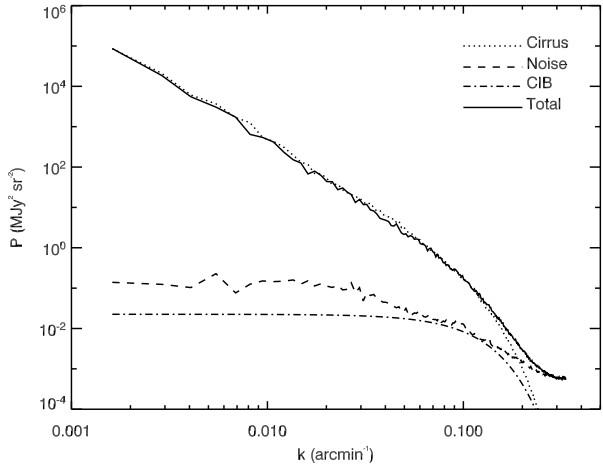}
\caption{\label{fig_powerspectrum} Power spectrum of
a typical IRIS map (point sources removed) with its associated noise power spectrum and the estimated CIB level (convolved 
by the IRAS beam). The power law at large scale (small $k$) is due to the Galactic
dust emission. The cutoff at $k\sim 0.1$~arcmin$^{-1}$ is due to the IRAS beam.}
\end{figure}

\section{The IRAS survey and IRIS}

\label{data}

In 1983 the Infrared Astronomical Satellite (IRAS) has made a survey
of 98\% of the sky in four bands : 12, 25, 60 and 100~\ump.
Since then this dataset has been used extensively in almost 
all area of astrophysics. In the early 90s the IRAS team
released the IRAS Sky Survey Atlas, a set of 430 fields with better calibration.
Each field is a $12.5^\circ \times12.5^\circ$ image with a pixel size of $1.5'\times1.5'$. 
Recently \cite{miville-deschenes2005a} reprocessed the ISSA maps
to correct for residual calibration defects and stripping. This new set
of ISSA plates, named IRIS, improves significantly the quality of the IRAS data
by lowering the noise level and improving the photometry of the four bands. 
Our analysis is based on the 100~\um IRIS maps, available in native $12.5^\circ \times12.5^\circ$
cartesian maps or in Healpix vector at http://www.cita.utoronto.ca/$\sim$mamd/IRIS/.

One important parameter of our analysis is the estimate of the instrumental 
noise contribution to the power spectrum. We used only the IRIS maps for which we could
compute the contribution of the noise. To do so we used the fact that 
the original ISSA plates are 
the combination of up to three HCON images. The power spectrum
of the noise can be estimated by taking the power spectrum of
the difference between two HCON images, as described
in more details by \cite{miville-deschenes2002b}. 

We selected only maps 
for which each pixel has been observed at least two times so that a noise map can be computed,
which represent 236 maps (out of 430).
Most maps of our sample have an average brightness $<I>$ lower than 20~MJy~sr$^{-1}$.
This is representative of the whole sky statistics, as seen in Fig.~\ref{fig_cumulativeHisto}
(see also Table~\ref{table:stat})
where the probability density function (PDF) and cumulative histogram of the 100~\um brightness (CIB and zodiacal light subtracted) 
over the whole sky is presented. This figure, together with Table~\ref{table:stat}, show that 90~\% of the sky
has a brightness lower or equal to 20~MJy~sr$^{-1}$ (or $N_H \lesssim 4\times10^{21}$~\cca
according to \cite{lagache2000a}). This simple statistics of 100~\um brightness reveals
that there is less than 2~\% of the sky with $N_H \lesssim 1\times10^{20}$~\ccap.

\begin{figure}
\includegraphics[width=\linewidth, draft=false]{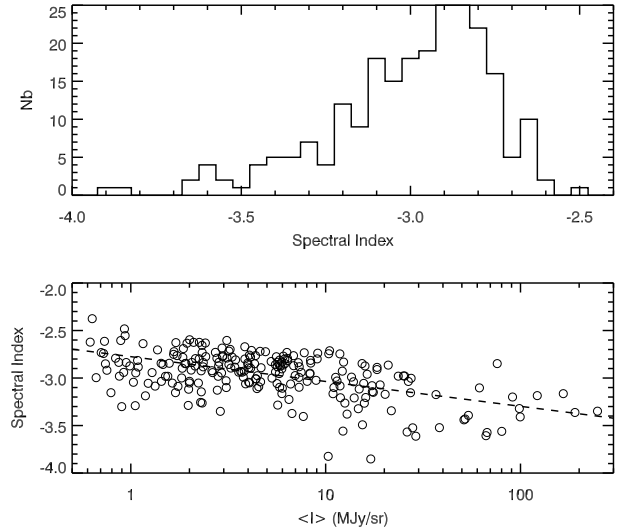}
\caption{\label{fig_spectralIndex} Spectral index $\gamma$ of the power spectrum for 
our sample. {\bf Top.} Histogram of the spectral index values. Median is -2.93, 
average is -2.96 and standard deviation is 0.3.
{\bf Bottom.} Spectral index as a function of average 100~\um brightness (CIB and zodiacal light subtracted).
}
\end{figure}

\section{Power spectrum analysis}

\label{power_spectrum_analysis}

The power spectrum has been extensively used in the analysis of image structure.
Working in Fourier (or spherical harmonics) space offers considerable advantages;
this formalism allows for the quadratic separation of uncorrelated sources 
in the image (e.g., noise and signal)
and deconvolution of the instrumental function.
In addition working in Fourier space facilitates the comparison with
numerical simulations of turbulent flows.

\subsection{Computation of the power spectrum}

The power spectrum of an image $f(x,y)$ of Fourier Transform 
$\tilde{f}(k_x,k_y)$ is computed from the
amplitude $A(k_x, k_y)$ defined as
\beq
A(k_x,k_y) = \tilde{f}(k_x,k_y)\tilde{f}^\star(k_x,k_y) = 
| \tilde{f}(k_x,k_y) |^2.
\eeq
The power spectrum $P(k)dk$ is an angular average 
of $A(k_x,k_y)$ between $k$ and $k+dk$ where $k = \sqrt{k_x^2 + k_y^2}$.
The method we use to compute the power spectrum is the one
described by \cite{miville-deschenes2002b}. 
To minimize edges effects in Fourier space we apodize the image, from which
the mean was removed, using a cosine tapper of 15 pixels wide.

\subsection{Contributions to the power spectrum}

The main goal of our analysis is to characterize the power
spectrum of the interstellar diffuse emission, but several astrophysical 
signals and artifacts may contribe to the power spectrum of the IRIS maps. 
The power spectrum $P(k)$ of the IRIS maps may be formalized by the following equation:
\beq
\label{eq_power_spectrum}
P(k) = B(k) (P_{ism} (k) + P_{sources} (k) + P_{cib}) + N(k)
\eeq
where $B(k)$ is the effective beam of the IRIS maps (which includes
the instrumental beam and the map making), $P_{ism} (k)$,  
$P_{sources} (k)$, $P_{cib} (k)$ and $N(k)$ are respectively the 
contributions of the interstellar medium, detected point sources, the 
unresolved CIB and the noise. All these contributions have to be estimated 
to characterize the interstellar component.

\begin{figure}
\includegraphics[width=\linewidth, draft=false]{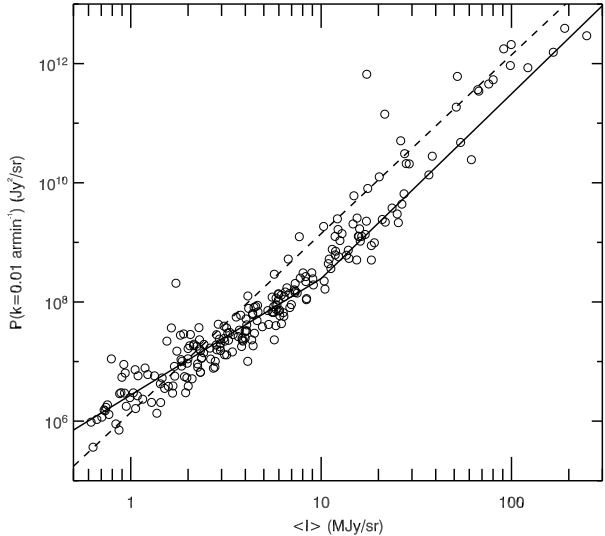}
\caption{\label{fig_normalisationPs} 
Normalisation $P(0.01)$ of the power spectrum at 0.01 arcmin$^{-1}$ of each map
as a function of its mean 100~\um brightness. 
The solid line is our fit to the data (see Equations~\ref{eq_normalisation_1} 
and \ref{eq_normalisation_2}). The dashed line is the relation given by \cite{gautier1992}.}
\end{figure}

A typical power spectrum of a faint region of the sky is shown in 
Fig.~\ref{fig_powerspectrum}.
At large scales (low $k$), the power spectrum follows a power law 
$A k^\gamma$; 
this is the signature of the cirrus cloud emission \cite[]{gautier1992}.
At smaller scales, the power spectrum flattens due to the combination
of the noise, point sources and the CIB. These three components
have a power spectrum signature that is flatter than the cirrus emission.

As stated in \S~\ref{data}, 
the power spectrum of the noise is estimated by taking the power spectrum of
the difference between two HCON images. An example of a noise map 
and its power spectrum is shown in Fig.~1 of \cite{miville-deschenes2002b}. 
The power spectrum of the noise is well described by a $k^{-1}$ 
power spectrum over most of the $k$ range.

To subtract the contribution of bright point sources 
we prefered to removed them directly in the IRIS maps
prior to compute the power spectrum. To do so 
we used the point source extraction method 
described by \cite{miville-deschenes2005a}.
For the CIB we assumed a flat power spectrum at the level determined
by  \cite{miville-deschenes2002b} ($5.8\times 10^3$~Jy$^2$~sr$^{-1}$ at 100~\ump).
Like \cite{miville-deschenes2002b} we made the assumption that the effective
beam of the IRIS maps is Gaussian with a FWHM of 4.3 arcminutes.


\subsection{Results}

To obtain the spectral index of the dust emission in each IRIS map, 
we computed the power spectrum of the point source subtracted map
and the power spectrum of the corresponding noise map.
We then subtracted the noise power spectrum, 
divide the result by the Gaussian PSF and remove the CIB contribution. 

Following what was done by \cite{gautier1992} we have fitted the power spectrum
of the Galactic emission in the 
range between $k=0.004$ and $k=0.08$~arcmin$^{-1}$ using a power law:
\begin{equation}
P(k) = P_0 \left(\frac{k}{k_0} \right)^\gamma
\end{equation}
where $P_0$ is the power spectrum value at $k_0=0.01$~arcmin$^{-1}$.
The power spectrum of the interstellar emission is in general very
well fitted by such a power law.

\begin{figure}
\includegraphics[width=\linewidth, draft=false]{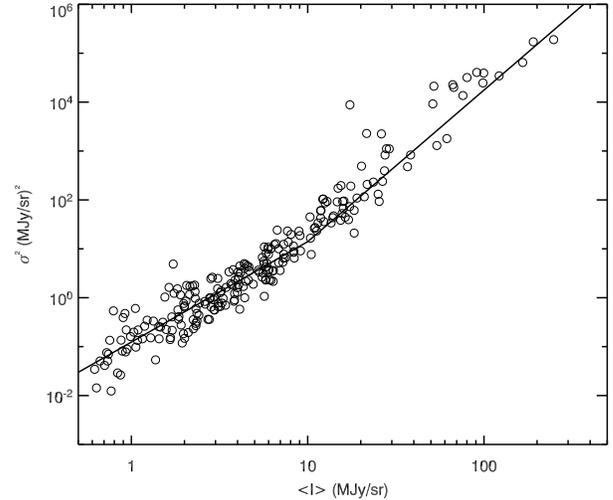}
\caption{\label{fig_variance_i} Standard deviation as a function of the average interstellar brightness
at 100~\um
for each map of our sample. The noise level of each map and the CIB fluctuation level 
(0.09~MJy~sr$^{-1}$ - see \cite{miville-deschenes2002b})
were removed quadratically from $\sigma$.
The solid line represents the two regimes given by equations~\ref{eq_sigmaL_1}
and \ref{eq_sigmaL_2}.}
\end{figure}

\subsubsection{Spectral index}

The compilation of the power spectrum spectral index measured on our sample of 236 maps is shown 
in Fig.~\ref{fig_spectralIndex}.
The most probable spectral index is $\gamma=-2.9\pm0.2$, in accordance with what was measured
by \cite{gautier1992}. On the other hand, contrary to \cite{gautier1992} who
did their statistical analysis on only four regions, our larger sample allowed us to 
highlight a significant variation of the spectral index with the average brightness of the maps.
Brighter regions on the sky tend to have a steeper power spectrum (see the lower panel of Fig.~\ref{fig_spectralIndex}).
The decrease of the spectral index with brightness can be approximated by:
\begin{equation}
\label{eq_gamma_I}
\gamma = -0.26 \log_{10}(<I>) - 2.77,
\end{equation}
where $<I>$ is the mean 100~\um brightness of the $12.5^\circ\times12.5^\circ$ map.

\begin{figure*}
\includegraphics[width=\linewidth, draft=false]{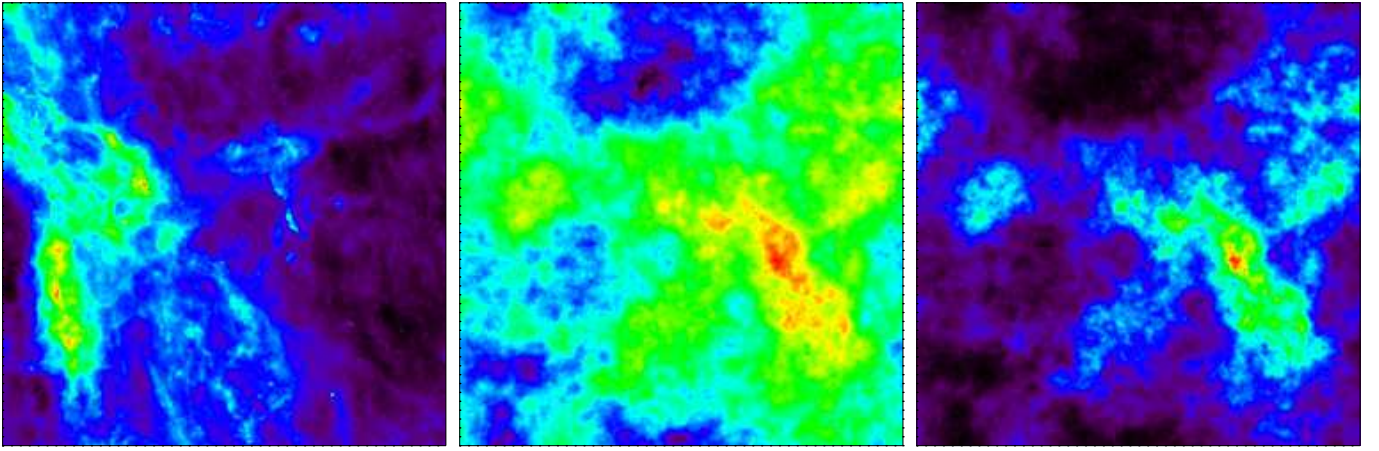}
\caption{\label{fig_image_simu} {\bf Left:} A typical IRIS map at 100~\ump. {
\bf Center:} A classical fBm map with same power spectrum as the IRIS map shown on the left.
This fBm has Gaussian brightness fluctuations at all scale and everywhere. 
{\bf Right:} A modified version of the classical fBm with only positive values and the same average,
standard deviation and skewness values as the IRIS map. The fBm has been modified in order to reproduce
the $\sigma(I) \propto <I>$ relation which results in stronger brightness fluctuations in brighter
regions.}
\end{figure*}

\subsubsection{Normalisation}

Regarding the normalization of the power spectrum \cite{gautier1992} estimated that
$P_0 = 1.4\times 10^6 <I>^3$, where $<I>$ is the mean 100~\um brightness at the 12.5$^\circ$ scale, 
given in MJy~sr$^{-1}$, and $P_0$ is in Jy$^2$~sr$^{-1}$. 
In Fig.~\ref{fig_normalisationPs} we present the variation of $P_0$ with $<I>$ for our sample, 
together with the relation of \cite{gautier1992} (dashed line).
The result of our analysis showed that the relation of \cite{gautier1992}
generally overestimates the fluctuation level at a given mean brightness.
This discrepancy could be partly attributed to the fact that we used better calibrated IRAS data 
compare to the early IRAS product used by \cite{gautier1992}.
On the other hand it is important to point out that 
the normalization relation given by \cite{gautier1992} is compatible
with the fact that they used only two faint ($<I>\sim 4$~MJy~sr$^{-1}$)
and two bright regions ($<I>\sim 30$~MJy~sr$^{-1}$).

Using a much larger sample than \cite{gautier1992} we found that the normalization of the power spectrum
is better described by two regimes (solid line in Fig.~\ref{fig_normalisationPs}):
on $\sim80$\% of the sky $P(0.01)$ is proportional to $<I>^{2.0\pm0.1}$ instead of $<I>^3$.
For $<I> <$~10~MJy~sr$^{-1}$ we find
\begin{equation}
\label{eq_normalisation_1}
P(0.01) = 2.7\times 10^6 <I>^{2.0\pm0.1}.
\end{equation}
and for $<I> \geq$~10~MJy~sr$^{-1}$ we find
\begin{equation}
\label{eq_normalisation_2}
P(0.01) = 2.0\times 10^5 <I>^{3.1\pm0.1}.
\end{equation}
A similar trend was also observed by \cite{jeong2005} using ISOPHOT observations (see their Fig.~12).

Another way of looking at the $P(0.01)-<I>$ relation is to plot the
standard deviation $\sigma_L^2$ of each map as a function of $<I>$ (see Fig.~\ref{fig_variance_i}).
Here we made sure to remove quadratically the contribution of the instrumental noise
and the CIB to each values of $\sigma_L^2$.
Like for the power spectrum normalisation, two regimes are apparent in the 
$\sigma_L^2$ vs $<I>$ relation with a transition around $<I>=$~10~MJy sr$^{-1}$. 
Separating the data sample in two gives the following fits.
for $<I> < 10$ MJy~sr$^{-1}$:
\begin{equation}
\label{eq_sigmaL_1}
\sigma_L^2 = 0.12<I>^{2.0\pm0.1} \mbox{(MJy$^2$ sr$^{-2}$)}
\end{equation}
and for $<I> \geq 10$ MJy~sr$^{-1}$
\begin{equation}
\label{eq_sigmaL_2}
\sigma_L^2 = 0.011 <I>^{3.1\pm0.1}  \mbox{(MJy$^2$ sr$^{-2}$)}
\end{equation}

\subsection{Interpretation}

\subsubsection{Spectral index and density structure}

Most spectral indexes measured here fall in the range between $\gamma=-3.6$ to $\gamma=-2.5$.
This is compatible with what was found by \cite{wright1998} 
($\gamma = -3$) in a power spectrum analysis of the DIRBE data at 60, 100, 140 and 240~\um on 
scales greater than 40 arcminutes.
Other studies also estimated the equivalent of the power spectrum spectral index
of the 100~\um emission using the area-perimeter relation. 
Following \cite{stutzki1998} there is a direct relation between the fractal dimension 
$D$ deduced from the area-perimeter relation ($p \propto a^{D/2}$) and the spectral
index $\gamma$ of the power spectrum ($P \propto k^{\gamma}$):
\begin{equation}
\gamma = 2D-6.
\end{equation}
\cite{bazell1988} and \cite{dickman1990} measured $D=1.25\pm0.05$ (corresponding to $\gamma = -3.5\pm0.1$) 
for relatively bright regions at 100~\um ($\sim10$~MJy~sr$^{-1}$). 
These values of $\gamma$ are rather high (steep power spectrum) 
but not incompatible with our analysis. In a similar analysis on fainter regions \cite{vogelaar1994}
measured $D=1.45\pm0.1$, corresponding to $\gamma=-3.1\pm0.2$ which is very similar
to our results. 

The interpretation of these results in terms of the density structure of the ISM
and the comparison with what is deduced from gas tracers 
can only be done by considering a model for the 100~\um emission. 
At this wavelength the interstellar emission is dominated by the grey body emission 
from big dust grains at thermal equilibrium with the radiation field \cite[]{desert1990}.
For a given model of the composition of dust grain \cite[]{desert1990,li2001} 
the conversion of 100~\um brightness to gas column density depends on the 
gas/dust mass ratio (known to be rather constant in the ISM) and on the big grain equilibrium
temperature. The big grain equilibrium temperature is related to the local radiation field strength 
and spectrum which depends on the presence or not of nearby heating sources and on the extinction.
Variation of the dust equilibrium temperature can also occur locally due to variation of
the grain structure which can affect their emissivity\footnote{Aggregation of small dust particles 
on bigger grains can increase significantly their area / mass ratio which leads to a more efficient cooling of the grains.
See \cite{stepnik2003} for a striking example of this effect.}.

In diffuse regions of the sky at high Galactic latitudes, far from star-forming regions, 
clouds are optically thin to stellar radiation, 
and the radiation field is uniform which result in very limited variations
of dust equilibrium temperature.
Localised variations of the dust grain temperature were observed in cirrus clouds \cite[]{bernard1999}, 
but overall several studies \cite[]{boulanger1988,boulanger1996} showed a strong correlation between
the 100~\um micron and the hydrogen column density which is in favor of
a rather uniform gas/dust ratio and limited variations of the dust temperature.
Based on these results we believe that the 100~\um micron can not be used as a
perfect surrogate for gas column density but overall it does not introduce 
a systematic bias in the determination of the column density power spectrum.
In this context, and considering that the power spectrum of the column density gives
directly the power spectrum of the density structure \cite[]{miville-deschenes2003b}, 
we consider that the $\gamma$ values measured here in the diffuse regions (for $<I>$ $<$ 10~MJy~sr$^{-1}$)
are typical of the spectral index of the density field in three dimensions in the solar neighborhood.

In bright regions ($<I>$ $>10$~MJy~sr$^{-1}$) the power spectrum is observed to be significantly steeper 
(Fig~\ref{fig_spectralIndex}). 
A steepening of the power spectrum with brightness was also found by \cite{kiss2003} 
on 90-200~\um ISOPHOT observations.
We observed that this steepening coincides with a departure from the $\sigma \propto <I>$ 
relation (see Fig.~\ref{fig_normalisationPs}) and a systematic increase of the skewness and kurtosis of the brightness 
fluctuations (see Fig.~\ref{fig_skewnessKurtosis}).
This variation of $\gamma$ with $<I>$ might reflect, at least in part, local variations of the density power spectrum.
It could also be attributed to the effect of gravity or anisotropic radiation fields 
which would both increase the large scale coherence (like in star forming regions or at Galactic scale in the plane) 
and therefore steepen the power spectrum of the 100~\um emission. The effect of extinction
in these parts of the sky will also induce important dust temperature variations which will affect
the power spectrum. These effects will certainly have an impact on the observed brightness fluctuations
and make the interpretration of the power spectrum difficult. 

To compare the results obtained here with statistical analysis of gas emission, 
it is important to point out that the 100~\um emission traces all the interstellar components: 
atomic, molecular and ionized gas. Gas observations (21 cm and CO) showed that
diffuse and dense gas have different density structure.
In general regions of cold and dense gas have stronger small scale fluctuations 
(and therefore a flatter density power spectrum
- see for instance \cite{stutzki1998} who measured $\gamma=-2.8$ on CO observations of the Polaris flare,
a dense cirrus cloud with a significant fraction of molecular gas) 
than regions of diffuse gas (\cite{miville-deschenes2003c} measured $\gamma=-3.6\pm0.2$ 
on 21 cm observations of a diffuse cirrus cloud). 
This behavior seems also in accordance with numerical simulations (see \cite{audit2005} for instance).

This could reflect the fact that molecular tracers like CO reveal only the density structure of
dense regions and miss the more diffuse and large scale structure of molecular clouds.
In that respect dust emission would be a more reliable tracer of the global density structure of
the ISM as the observed spectral index at 100~\um is a weighted mean of the various contributions 
from dense and diffuse media on the line of sight. 
One striking example of that comes from the direct comparison of infrared and 21 cm emission in high-latitude clouds
where dust emission usually shows stronger brightness 
fluctuations at small scale than \hip, this being attributed to the presence of localized molecular regions
(see for instance \cite{joncas1992}).

\subsubsection{The $\sigma \propto <I>$ relation}

Here we would like to propose one interpretation for the scaling of
the power spectrum normalization $P(0.01)$ (or equivalently the brightness 
standard deviation $\sigma_L$) with average brightness $<I>$ 
(equations~\ref{eq_normalisation_1} and \ref{eq_sigmaL_1}). 

Lets consider a three-dimensional scalar field $\epsilon$, which could be a 
dust emissivity field, of size $L\times L$ on the plane of the sky and of depth $H$. 
In the case of constant dust temperature and neglecting opacity
effects, the projection of this 3D field on 2D would correspond to a dust emission map:
\begin{equation}
I(x,y) = \int_0^H \epsilon(x,y,z) \, dz.
\end{equation}
What would be the relation between the average and standard deviation of the projected brightness 
for such a field~?
The average brightness of the dust map is simply
\begin{equation}
\label{eq_I_vs_H}
<I> = <\epsilon>H
\end{equation}
where $<\epsilon>$ is the volume averaged of $\epsilon$.

On the other hand to compute the standard deviation of the dust map ($\sigma_I$)
one should consider that brightness fluctuations are added quadratically 
along the line of sight. Each slice $dz$ of the cube contributes $\sigma_\epsilon^2 dz$
to $\sigma_I^2$:
\begin{equation}
\label{eq_sigma_cartesian}
\sigma^2_I = \int_0^H \sigma_\epsilon^2 \, dz \, = \sigma_\epsilon^2 H.
\end{equation}
Combined with Eq.~\ref{eq_I_vs_H} this leads to 
\begin{equation}
\sigma_I^2 \propto <I>
\end{equation}
which is not what is observed.

Cartesian coordinates used in the previous demonstration is
in fact not a realistic representation when it comes to estimate $\sigma_I$. 
In fact, with the increase of the depth of the line of sight $z$, 
a given angular scale $\theta$
on the sky corresponds to increasing physical size $l$ : $\tan(\theta)=l/z$.

In the integral of Eq.~\ref{eq_sigma_cartesian} we made the assumption that $\sigma_\epsilon$
is the standard deviation at scale $L$. In fact we should have taken into account
that $\sigma_\epsilon$ varies with $z$. 
For a power spectrum following a power law ($P(k) \propto k^\gamma$) the 
variance of brightness fluctuations as a function 
of scale $l$ is \cite[]{brunt2002a}:
\begin{equation}
\label{eq_sigma_l}
\sigma_l \propto l^{-\gamma/2-1}.
\end{equation}

Therefore we can rewrite Eq.~\ref{eq_sigma_cartesian}
\begin{equation}
\sigma^2_I(\theta) \propto \int_0^H l^{-\gamma-2} \, dz = \int_0^H (\tan(\theta) z)^{-\gamma-2} \, dz 
\end{equation}
which leads to
\begin{eqnarray}
\sigma^2_I(\theta) & \propto & \tan(\theta)^{-\gamma-2} \, H^{-\gamma-1}\nonumber\\ 
 & \propto & \tan(\theta)^{-\gamma-2} \, <I>^{-\gamma-1}.
\end{eqnarray}
There are two important results to extract from the previous equation. First
the fact that we observed structure of increasing physical size with $z$
at a given angular scale $\theta$ does not modify the slope of the power law for $\theta\lesssim 25^\circ$:
both $\sigma_I(\theta)$ and $\sigma_\epsilon(l)$ have the same spectral index $(\gamma-2)/2$.
Secondly the dependence of $\sigma_I$ on $<I>$ is now related to the spectral index $\gamma$.
For a typical value of $\gamma=-3$ we obtain $\sigma_I^2 \propto <I>^2$ which is exactly
the observed relation for $<I> < 10$~MJy~sr$^{-1}$.

For larger $<I>$ the departure from a simple $\sigma_I^2 \propto <I>^2$ can be explained
partly because the spectral index $\gamma$ decreases but most of all it is important to notice
that brightness fluctuations become highly non-Gaussian (high skewness and kurtosis - 
see Fig.~\ref{fig_skewnessKurtosis}). In this regime the standard deviation becomes 
significantly affected by non-Gaussian fluctuations which could be related to localized
star forming regions in the image.

\begin{figure}
\includegraphics[width=\linewidth, draft=false]{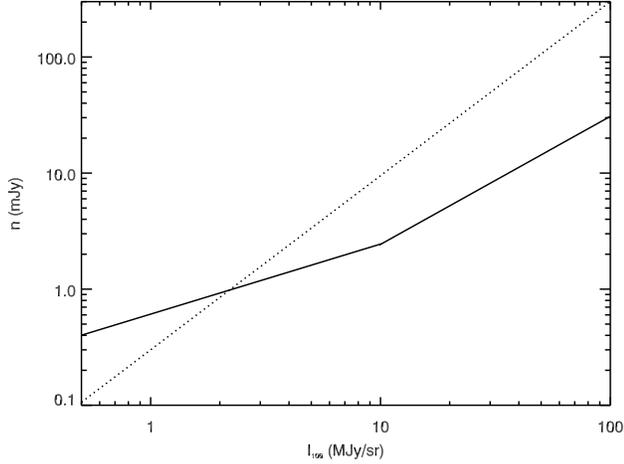}
\caption{\label{fig:compare_helou1990} Cirrus noise at 100~\um for a 1m diffraction limited telescope 
(FHWM=33 arcseconds) as a function of $I_{100}$ brightness. 
Solid line is our estimate and dotted line is the estimate of \cite{helou1990}.}
\end{figure}


\section{Cirrus noise}

In this section we use the properties of the power spectrum of 100~\um emission
to estimate the level of cirrus confusion noise of dust emission.

\label{cirrus_noise}

\subsection{Brightness fluctuations}

To estimate the level of fluctuations as a function of scale and brightness
it is convenient to use the real-space representation given by Eq.~\ref{eq_sigma_l}, as opposed 
to a Fourier one.
Using the value of $\sigma_l$ measured at $l=L=12.5^\circ$ (see equations~\ref{eq_sigmaL_1} and \ref{eq_sigmaL_2})
we can normalize Eq.~\ref{eq_sigma_l} and estimate 
the level of brightness fluctuation of interstellar dust (in MJy~sr$^{-1}$) at scale $l$ (in degree)
and wavelength $\lambda$. For the two regimes identified in this study 
(lower and higher than $<I_{100}>$=10~MJy~sr$^{-1}$) we have the following relations:
\begin{equation}
\label{eq:noise_bright1}
\sigma_{l,\lambda}^{low} = 0.35 <I_{100}> \frac{I_\lambda}{I_{100}}
\left( \frac{l}{12.5^\circ}\right)^{-\gamma/2-1} 
\end{equation}
and
\begin{equation}
\label{eq:noise_bright2}
\sigma_{l,\lambda}^{high} = 0.10 <I_{100}>^{1.5} \frac{I_\lambda}{I_{100}}
\left( \frac{l}{12.5^\circ}\right)^{-\gamma/2-1}.
\end{equation}
where $<I_{100}>$ is in MJy~sr$^{-1}$, $\gamma$ is given by Eq.~\ref{eq_gamma_I}
and $I_\lambda/I_{100}$ is an estimate of the dust brightness ratio at wavelength $\lambda$
and 100~\ump. Such ratio can be estimated using a grey body model like the one of \cite{desert1990}
\begin{equation}
\frac{I_\lambda}{I_{100}} =  \frac{B_\nu(T_d)}{B_{100}(T_d)} \left( \frac{\lambda}{100 \, \mu m} \right)^{-\beta}
\end{equation}
where $B_\nu(T_d)$ is the Planck function at grain temperature $T_d$ and $\beta$ is the dust emissivity index.
Typical values in the diffuse ISM for the dust temperature $T_d$ and emissivity index $\beta$ are $T_d=17.5$~K and $\beta=2$.
In the sub-millimeter and millimeter ranges the dust emission departs from this simple model and
one might want to use a combination of two dust components (see \cite{finkbeiner1999} for an example of such model).

\begin{figure}
\includegraphics[width=\linewidth, draft=false]{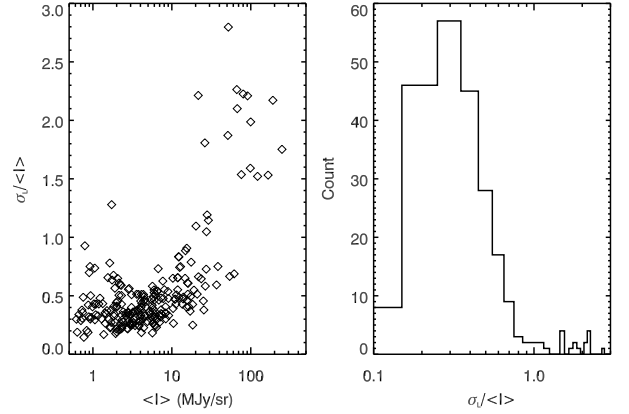}
\caption{\label{fig_contrast} Contrast of interstellar emission: the standard deviation 
$\sigma_L$ is the standard deviation of interstellar fluctuations at the $12^\circ$ scale.
{\bf Left: } Contrast values as a function of average brightness. 
{\bf Right: } Histogram of contrast values. According to Eq.~\ref{eq_sigmaL_1}, the contrast is close to 
constant for $<I>$ lower than 10~MJy~sr$^{-1}$. Outliers at low brightness corresponds to regions
where there are bright structures on a very low level background, like the Magellanic Clouds.}
\end{figure}

\begin{figure*}
\includegraphics[width=\linewidth, draft=false]{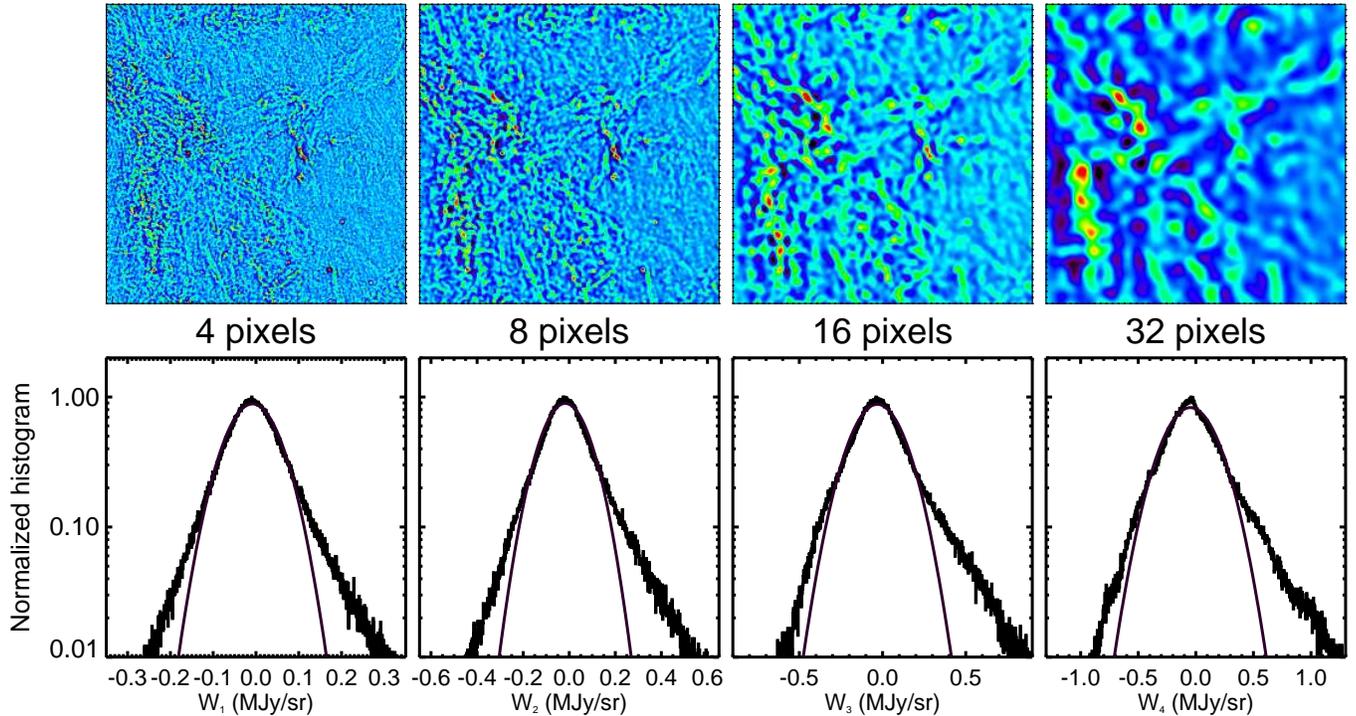}
\caption{\label{fig_wavelet_iris} Wavelet decomposition of the IRIS map shown in 
Fig.~\ref{fig_image_simu}-left.
{\bf Top:} wavelet coefficients map at scales 4, 8, 16 and 32 pixels (1 pixel=1.5'). 
{\bf Bottom: } Histogram of the wavelet coefficients in linear-log.  
A Gaussian fit to the histogram is superposed highlighting the non-Gaussian behavior.}
\end{figure*}

\subsection{Noise per beam}

Equations~\ref{eq:noise_bright1} and \ref{eq:noise_bright2} give the surface brightness fluctuation level of dust
at any scale, brightness and wavelength. One useful specific case to consider is the contribution 
of cirrus noise at the scale of an instrument beam, to estimate the effective point source detection level for instance.
Following \cite{helou1990} and \cite{gautier1992} we consider the cirrus noise level at a scale two times the 
beam size $b$ (i.e., $l=2b$, where $b$ is the beam FWHM).
This noise level (in mJy/beam) at wavelength $\lambda$ is simply
\begin{equation}
n_\lambda = 10^9 \, \sigma_{2b,\lambda} \, \Omega
\end{equation}
where $\Omega$ is the beam in steradian
\begin{equation}
\Omega = \pi \left(\frac{b}{2}\right)^2 \left(\frac{\pi}{180}\right)^2
\end{equation}
and $b$ is in degrees.
The cirrus noise in mJy/beam is then given by
\begin{equation}
n_{\lambda}^{low} = 3.3\times 10^6 <I_{100}> \frac{I_\lambda}{I_{100}} \left(0.16 \, b\right)^{-\gamma/2+1}
\end{equation}
and
\begin{equation}
n_{\lambda}^{high} = 1.0\times 10^6 <I_{100}>^{1.5} \frac{I_\lambda}{I_{100}} \left(0.16 \, b\right)^{-\gamma/2+1}
\end{equation}
where $\gamma$ is still given by Eq.~\ref{eq_gamma_I}.

These relations can be compared with the prescription of \cite{helou1990} often used
to estimate cirrus confusion noise. These authors considered the situation where the beam of the instrument
is given by the diffraction limit of a telescope of diameter $D$. In this case $b= (1.6\lambda / D) (180/\pi)$.

The comparison between our estimate of the cirrus noise and the one given by \cite{helou1990} for a 1m telescope
at 100~\um is shown in Fig~\ref{fig:compare_helou1990}.
The difference is important in several aspects. First at low brightness the slopes of $n(I_{100})$ are very different, 
due to the $\sigma \propto <I>$ regime revealed in our study. The change of slope has the implication
that our estimate of the cirrus noise in low brightness regions is much higher. This is also due to 
the flattening of the power spectrum index $\gamma$ at low brightness. 
On the other hand our cirrus noise estimate is much lower than the estimate of \cite{helou1990}
at large brightness. This effect is caused by the fact that we observe steeper power spectra
in bright regions. Therefore the brightness fluctuations are much smaller when extrapolated to small scale.

\subsection{Limitation}

The prescriptions given here to estimate the level of cirrus noise is only indicative. 
It would be strictly correct for a Gaussian field which is not the case of interstellar emission.
As it will be described in more details in the next section, dust emission
has non-Gaussian brightness fluctuations at all scales. 
For the sake of data interpretation, confusion estimate
or tests of component separation algorithms (including point source extraction)
it is useful to produce realistic simulations of dust emission maps
with proper non-Gaussian properties. It is the topic of the next section.

\begin{figure*}
\includegraphics[width=\linewidth, draft=false]{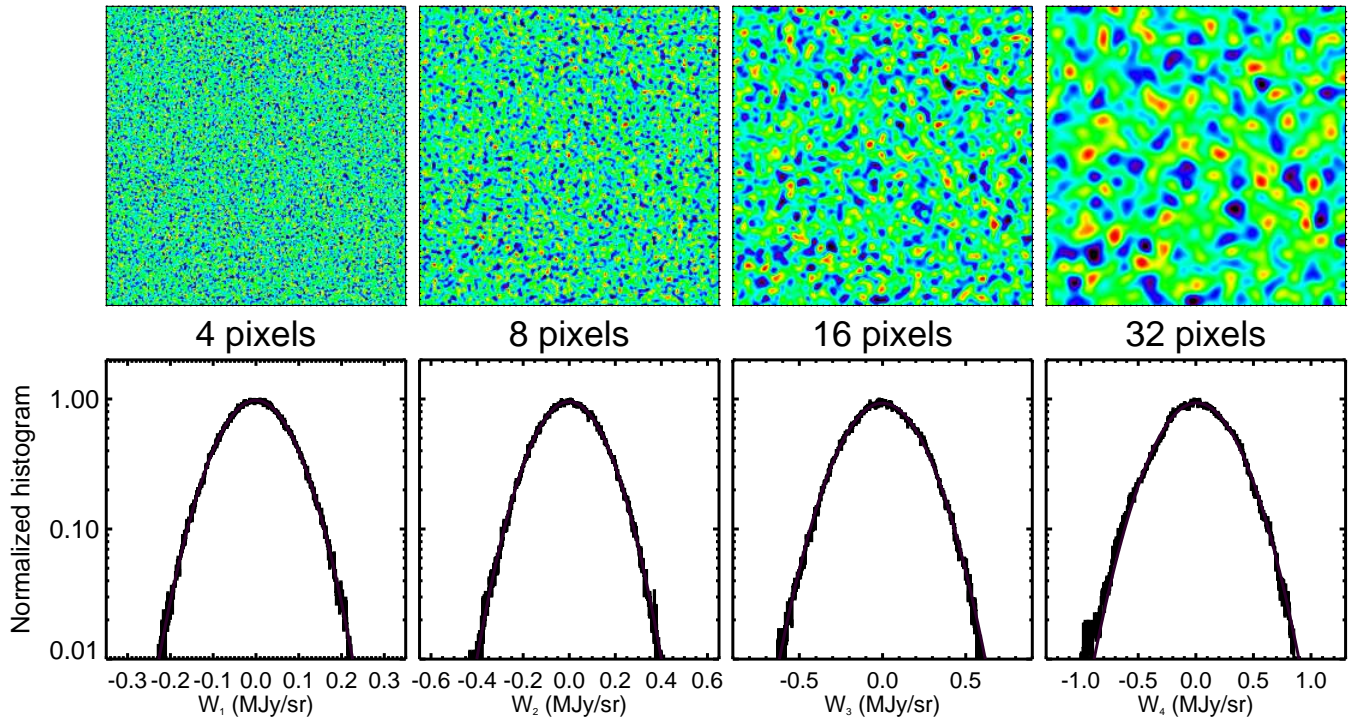}
\caption{\label{fig_wavelet_fbm0} Wavelet decomposition of the classical fBm map shown 
in Fig.~\ref{fig_image_simu}-center. See Fig.~\ref{fig_wavelet_iris} for details.}
\end{figure*}


\section{Construction of artificial dust emission maps}

\label{artificial_maps}

\subsection{Fractional Brownian motion}

Fractional Brownian motions (fBm, also known as Gaussian random fields) are often
used to simulate interstellar emission \cite[e.g.,][]{stutzki1998,miville-deschenes2003b}. 
By construction such objects can reproduce the power spectrum of any image, 
with the limitation that its phase is random. A comparison of a typical IRIS map and a 
fBm with the same power spectrum is given in Fig.~\ref{fig_image_simu}.
The fBm reproduces well the self-similar structure of the interstellar emission.
On the other hand the fBm seems smoother and less contrasted than the observation.

Because of the Gaussian nature of their fluctuations, fBms with positive values only 
can't have a large contrast.
If we define the contrast of a map $I$ as $C_I = \sigma(I)/<I>$, positive fBms are restricted
to $C_I \lesssim 1/3$ as the average $<I>$ needs to be greater than $3\sigma$ to have
only positive values\footnote{or equivalently that 99~\% of the values are 
between $<I>-3\sigma$ and $<I>+3\sigma$.}.
This can be compared with the contrast of real IRIS map given in Fig.~\ref{fig_contrast}, 
where we used $\sigma(I)=\sigma_L$ (i.e., the standard deviation at a scale of 12.5$^\circ$). 
The median contrast is 0.3 which indicate the limitation of the use of fBms used to simulate realistic
infrared dust maps. In addition one would notice that the contrast increases significantly
with brightness and especially for $<I>$ greater than 10~MJy~sr$^{-1}$, in accordance with equations~\ref{eq_sigmaL_1} 
and \ref{eq_sigmaL_2}.
In addition, $C_I$ will increase with scale as $\sigma(I)$ depends on scale 
(according to Eq.~\ref{eq_sigma_l}) 
but not $<I>$ (at high Galactic latitude). 
Based only on the contrast, the use of fBm becomes limited to small (a few degrees) and faint regions of the sky.

\subsection{Wavelet decomposition}

Apart from the global contrast limitation, there is a fundamental difference between
fBms and observations which is related to the non-Gaussian properties of the interstellar
emission.
The brightness fluctuations seen in infrared dust maps show contrasted structures, often filaments,
that reflect their non-Gaussian nature. Using a wavelet decomposition
of 100~$\mu$m maps, \cite{jewell2001} showed clearly that, contrary to random-phase realisations, 
the histograms of brightness fluctuations at a given scale are highly non-Gaussian.
\cite{abergel1996a} and  \cite{aghanim2003} gave also striking examples of that.
Wavelet transforms are powerful
tools to study the statistical moments higher than two and estimate the non-Gaussian
properties of images \cite[]{aghanim2003}.
They complement the power spectrum analysis which gives only the
variation of the second moment - the standard deviation - with scale. 

To illustrate that we present in figures~\ref{fig_wavelet_iris} and \ref{fig_wavelet_fbm0} the
wavelet decomposition obtained using the ``a trou'' algorithm \cite[]{starck1998}
for the IRIS and fBm maps of Fig.~\ref{fig_image_simu}. 
As expected the distribution of brightness fluctuations at a given scale are Gaussian
for fBms (see Fig.~\ref{fig_wavelet_fbm0}). On the other hand 
the distribution of wavelet coefficients of an IRIS map (see Fig.~\ref{fig_wavelet_iris}) follows an 
asymmetrical distribution with exponential wings, which results in significant
skewness and kurtosis.

\begin{figure}
\includegraphics[width=\linewidth, draft=false]{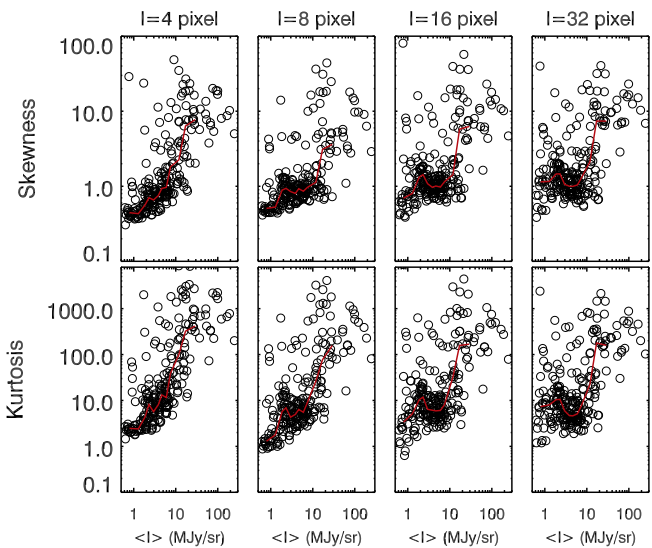}
\caption{\label{fig_skewnessKurtosis} Skewness (top) and Kurtosis (bottom)
of the wavelet coefficients from l=4 to 32 pixels for each IRIS map of our sample. 
The solid red line is the median in a bin 20 values sorted by increasing $<I>$.}
\end{figure}

In Fig.~\ref{fig_skewnessKurtosis} are compiled 
the skewness and kurtosis of the wavelet coefficients found for all 236 maps of our sample, 
for scales from $l=4$~pixels (12 arcmin) to 32 pixels (48 arcmin).
There is a not a strong correlation of the skewness and kurtosis with brightness, unlike
for the standard deviation (Fig.~\ref{fig_variance_i}).
There is a slight increase of skewness and kurtosis at small scale and low brightness
that should be attributed to noise and CIB
\footnote{Both processes being more Gaussian than the interstellar emission they have the effect of lowering 
the skewness and kurtosis mostly. This effect is more important at small scale and low brightness where 
the noise and CIB fluctuations are stronger with respect to interstellar fluctuations.}.
On the other hand one would note a sharp increase of the skewness and kurtosis 
for $<I> > 10$~MJy~sr$^{-1}$. 

Finally, apart from the global difference of the wavelet coefficient distribution between fBms and
observations, one would notice that brightness fluctuations at a given scale in the IRIS map
are generally greater in bright regions of the map, contrary to the fBm where the amplitude of 
fluctuations at a given scale is independent of position. 
This just reflects the fact that brightness fluctuations are generally stronger in bright region,
in accordance with equations~\ref{eq_sigmaL_1} and \ref{eq_sigmaL_2}. 
By construction fBms do not behave like that; the amplitude of fluctuations is uniform and independant 
of local variations of the average brightness.

\begin{figure*}
\includegraphics[width=\linewidth, draft=false]{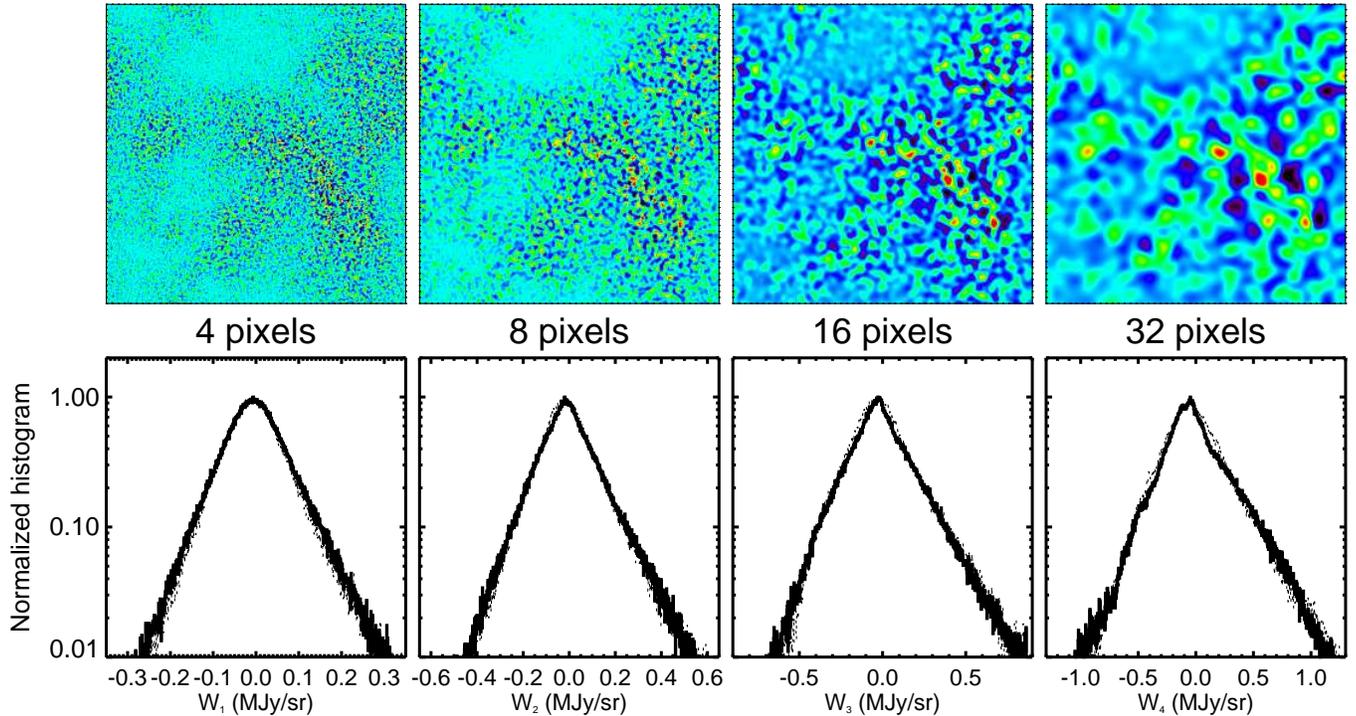}
\caption{\label{fig_wavelet_fbm1} Wavelet decomposition of the modified fBm map shown 
in Fig.~\ref{fig_image_simu}-right. See Fig.~\ref{fig_wavelet_iris} for details. The histogram
of the wavelet coefficients of the IRIS map are over plotted (dots) showing how well the
modified fBm reproduces the observed statistics.}
\end{figure*}

\subsection{Non-Gaussian fBm}

In this section we propose a method to modify fBms such that they better reproduce
the statistical properties of the infrared emission. We want 
to produce simulations of dust emission maps that would satisfy a given number of assumptions:
\begin{enumerate}
\item Positive values only
\item Power spectrum following a power law : $P(k) = Ak^\gamma$
\item Greater brightness fluctuations in bright regions (following $\sigma \propto <I>$).
\end{enumerate}

Several attempts have been made to generate non-Gaussian maps, especially to produce
non-Gaussian realizations of the Cosmic Microwave Background (CMB) \cite[]{vio2001,vio2002,rocha2005}.
In general these methods involve a modification of a Gaussian realization but they don't
comply with all our requirements. Especially in the context of the CMB there is no need
to produce maps with only positive values and to
have stronger brightness fluctuations in bright regions.

Here is how we proceeded to construct a non-Gaussian fBm map 
with the same statistical properties as an
IRIS map $I$ and with only positive values.
The method presented here is similar to the one 
used by \cite{elmegreen2002,brunt2002a}.
First we generate a classical fBm $F$ with same standard deviation as map $I$. 
We add an offset to force all values to be positives
and to be as close as possible to the average of $I$.
We then create a modified fBm $F'$ such that:
\begin{equation}
F' = A \, F^\psi \, + C
\end{equation}
and such that $F'$ has the same average, standard deviation and skewness than $I$
and only positive values. The parameters $A$, $\psi$ and $C$ are thus constrained iteratively 
so to match the first three statistical moments of the map $I$.
Such a modified fBm is shown in Fig.~\ref{fig_image_simu} and its wavelet decomposition
in Fig.~\ref{fig_wavelet_fbm1}. The effect of the exponent $\psi$ is to produce
greater fluctuations in bright regions, in accordance with the observations. The
wavelet decomposition is quite illustrative in that respect. In particular the
histograms of wavelet coefficients are almost impossible to discriminate from the
ones of the IRIS map. 

\begin{figure}
\includegraphics[width=\linewidth, draft=false]{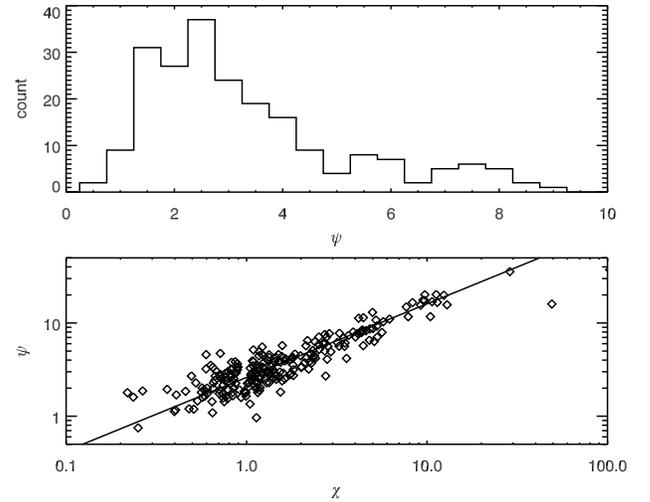}
\caption{\label{fig_alpha_skew} {\bf Top:} Histogram of all the exponent $\psi$ 
used to simulate realistic fBm for each IRIS map of our sample. {\bf Bottom:} 
Exponent $\psi$ as a function of the skewness $\chi$ of each IRIS map.
The straight line is $\psi = 2.6 \, \chi^{0.8}$.}
\end{figure}

The key parameter in this transformation is $\psi$ which
control the amount of non-Gaussianity introduced in the map. 
We did such a simulation for each of the 236 maps of our sample and the results are summarized
in Fig.~\ref{fig_alpha_skew}. One would note 
the scaling of $\psi$ with skewness $\chi$, which clearly indicate the impact
of that parameter on the non-Gaussianity of the map. The median value of $\psi$
is $\sim 2$ (see Fig.~\ref{fig_alpha_skew}). 

Even though it reproduces very well several statistical properties of the observed emission, 
it is important to note that one important limitation of this method
is that it produces only isotropic fluctuations and failed to reproduce to filamentary structure
of the ISM.

\subsection{Extend an image to smaller scales}

{
To estimate the capabilities and performances of some instruments at observing 
diffuse interstellar emission it is often needed to extrapolate low resolution 
observations to smaller angular scales. One example of that would be the estimate 
of the diffuse emission structure that will be observed at a scale of 8~arcsec by Herschel-PACS 
given the IRAS data at 5 arcmin resolution.
In this context it is useful to produce constrained realisations of the interstellar diffuse
emission based on low resolution observations. The statistical analysis presented here
provides a theoretical basis ($\sigma \propto <I>$) on which to add small scales using 
larger scale informations.

Given a low resolution map $I_0$ characterized by a beam $B_0$, a higher resolution $I_1$
of beam $B_1$ can be computed using the following method:
\begin{equation}
I_1 = I_0 + A \, (B_1\otimes F - B_0\otimes F) \, I_0^{\psi} + C
\end{equation}
where $F$ is a classical positive fBm map, 
and $A$, $\psi$ and $C$ are determined as described in the previous section. 
In the previous equation the difference $(B_1\otimes F-B_0\otimes F)$ extends the 
structure up to the higher resolution beam $B_1$.
The modulation of the small scale fluctuations by $I_0^{\psi}$ assures 
that they follow the $\sigma \propto <I>$ relation. 
}


\section{Conclusion}

In this paper we presented an analysis of the power spectrum and wavelet decomposition
of the IRIS/IRAS 100~\um emission over 55~\% of the sky.
The main goals of this work were 1) to review and extend the study of \cite{gautier1992} 
using better calibrated IRAS maps, estimates of the noise and CIB contributions
and a larger sample, 2) provide a more precise prescription for cirrus noise and 3) 
suggest a technique to simulate dust emission map with proper statistical properties.

We found an average spectral index ($<\gamma> = -2.9\pm0.2$) compatible with \cite{gautier1992} ($\gamma= -3$)
but with a significant variation from $\gamma=-3.6$ to $\gamma=-2.5$.
Considering that 100~\um emission is a relatively reliable tracer of column density in faint regions, 
these values of $\gamma$ should be representative of the spectral index of density in three dimensions
in the local interstellar medium. The comparison with other tracers leads to the conclusion
that there is most probably a significant contribution from cold gas 
to the brightness fluctuations observed.
We also found a slight variation of $\gamma$ with $<I>$ which could be explained by the
impact of gravity or spatial variations of dust temperature in star forming regions.

We also found that the amplitude of the brightness fluctuations were
generally overestimated by \cite{gautier1992}. In regions with a 100~\um average brightness $<I>$ 
lower than 10~MJy~sr$^{-1}$ the brightness fluctuation level is proportional to $<I>$ and 
not $<I>^{1.5}$ as stated by \cite{gautier1992}. 
We showed that this behavior can be explained by the fact that
the brightness fluctuation level observed at a given angular size on the sky
is the sum of fluctuations of increasing amplitude with distance.

This detailed description of the power spectrum properties of the 100~\um emission
allowed us to determine a new prescription of the cirrus confusion noise in the 
far-infrared and sub-millimeter as a function of column density and scale.
On the other hand we stressed that this cirrus noise estimate relies on the hypothesis of Gaussian
fluctuations, which is clearly not the case for interstellar emission. 
In that context we proposed a method to
modify Gaussian random fields such that it reproduces the power spectrum but also
the level of non-Gaussianity observed which is related to  
the fact that bright regions have stronger brightness fluctuations than faint ones.
Such images could be used to tests component separation algorithms (including point source extraction)
that have to deal with non-Gaussian components.
The main limitation of the technique we propose is that it does not reproduce the obvious filamentarity
seen in observations. 

\vspace*{1cm}

Some of the results in this paper have been obtained using the HEALPix package
\cite[]{gorski2005}. This work was supported by the Canadian Space Agency.
It is a pleasure to thank J. Richard Bond for enlightening discussions.


\begin{thebibliography}{Miville-Desch\^enes \& Lagache(2005)}

\bibitem[Abergel et~al.(1996)]{abergel1996a}
Abergel, A., Boulanger, F., Delouis, J.~M., Dudziak, G. \& Steindling, S.
\newblock 1996, \AaA, 309, 245.

\bibitem[{Aghanim} et~al.(2003)]{aghanim2003}
{Aghanim}, N., {Kunz}, M., {Castro}, P.~G. \& {Forni}, O.
\newblock 2003, \AaA, 406, 797--816.

\bibitem[Audit \& Hennebelle(2005)]{audit2005}
Audit, E. \& Hennebelle, P.
\newblock 2005, \AaA, 433, 1--13.

\bibitem[Bazell \& Desert(1988)]{bazell1988}
Bazell, D. \& Desert, F.~X.
\newblock 1988, \ApJ, 333, 353.

\bibitem[Bensch et~al.(2001)]{bensch2001}
Bensch, F., Stutzki, J. \& Ossenkopf, V.
\newblock 2001, \AaA, 366, 636.

\bibitem[Bernard et~al.(1999)]{bernard1999}
Bernard, J.~P., Abergel, A., Ristorcelli, I., Pajot, F., Torre, J.~P.,
  Boulanger, F., Giard, M., Lagache, G., Serra, G., Lamarre, J.~M., Puget,
  J.~L., Lepeintre, F. \& Cambr\'esy, L.
\newblock 1999, \AaA, 347, 640.

\bibitem[Boulanger \& P\'erault(1988)]{boulanger1988}
Boulanger, F. \& P\'erault, M.
\newblock 1988, \ApJ, 330, 964.

\bibitem[Boulanger et~al.(1996)]{boulanger1996}
Boulanger, F., Abergel, A., Bernard, J.~P., Burton, W.~B., Desert, F.~X.,
  Hartmann, D., Lagache, G. \& Puget, J.~L.
\newblock 1996, \AaA, 312, 256.

\bibitem[Brunt \& Heyer(2002)]{brunt2002a}
Brunt, C.~M. \& Heyer, M.~H.
\newblock 2002, \ApJ, 566, 276.

\bibitem[{Brunt}(2003)]{brunt2003}
{Brunt}, C.~M.
\newblock 2003, \ApJ, 583, 280.

\bibitem[Crovisier \& Dickey(1983)]{crovisier1983}
Crovisier, J. \& Dickey, J.~M.
\newblock 1983, \AaA, 122, 282.

\bibitem[Desert et~al.(1990)]{desert1990}
Desert, F.~X., Boulanger, F. \& Puget, J.~L.
\newblock 1990, \AaA, 237, 215.

\bibitem[Dickey et~al.(2001)]{dickey2001}
Dickey, J.~M., McClure-Griffiths, N.~M., Stanimirovic, S., Gaensler, B.~M. \&
  Green, A.~J.
\newblock 2001, \ApJ, 561, 264.

\bibitem[Dickman et~al.(1990)]{dickman1990}
Dickman, R.~L., Margulis, M. \& Horvath, M.~A.
\newblock 1990, \ApJ, 365, 586.

\bibitem[Elmegreen et~al.(2001)]{elmegreen2001}
Elmegreen, B.~G., Kim, S. \& Staveley-Smith, L.
\newblock 2001, \ApJ, 548, 749.

\bibitem[{Elmegreen}(2002)]{elmegreen2002}
{Elmegreen}, B.~G.
\newblock 2002, \ApJ, 564, 773.

\bibitem[Falgarone et~al.(1991)]{falgarone1991}
Falgarone, E., Phillips, T.~G. \& Walker, C.~K.
\newblock 1991, \ApJ, 378, 186.

\bibitem[Finkbeiner et~al.(1999)]{finkbeiner1999}
Finkbeiner, D.~P., Davis, M. \& Schlegel, D.~J.
\newblock 1999, \ApJ, 524, 867.

\bibitem[Gautier et~al.(1992)]{gautier1992}
Gautier, T. N.~I., Boulanger, F., Perault, M. \& Puget, J.~L.
\newblock 1992, \AJ, 103, 1313.

\bibitem[Goldman(2000)]{goldman2000}
Goldman, I.
\newblock 2000, \ApJ, 541, 701.

\bibitem[{Gorski} et~al.(2005)]{gorski2005}
{Gorski}, K.~M., {Hivon}, E., {Banday}, A.~J., {Wandelt}, B.~D., {Hansen},
  F.~K., {Reinecke}, M. \& {Bartelmann}, M.
\newblock 2005, \ApJ, 622, 759--771.

\bibitem[Green(1993)]{green1993}
Green, D.~A.
\newblock 1993, \MNRAS, 262, 327.

\bibitem[{Hauser} et~al.(1998)]{hauser1998}
{Hauser}, M.~G., {Arendt}, R.~G., {Kelsall}, T., {Dwek}, E., {Odegard}, N.,
  {Weiland}, J.~L., {Freudenreich}, H.~T., {Reach}, W.~T., {Silverberg}, R.~F.,
  {Moseley}, S.~H., {Pei}, Y.~C., {Lubin}, P., {Mather}, J.~C., {Shafer},
  R.~A., {Smoot}, G.~F., {Weiss}, R., {Wilkinson}, D.~T. \& {Wright}, E.~L.
\newblock 1998, \ApJ, 508, 25--43.

\bibitem[{Helou} \& {Beichman}(1990)]{helou1990}
{Helou}, G. \& {Beichman}, C.~A.
\newblock The confusion limits to the sensitivity of submillimeter telescopes.
\newblock In ESA, editor, {\em From Ground-Based to Space-Borne Sub-mm
  Astronomy}, page 117, 1990.

\bibitem[Hobson(1992)]{hobson1992}
Hobson, M.~P.
\newblock 1992, \MNRAS, 256, 457.

\bibitem[{Ingalls} et~al.(2004)]{ingalls2004}
{Ingalls}, J.~G., {Miville-Deschênes}, M.~A., {Reach}, W.~T.,
  {Noriega-Crespo}, A., {Carey}, S.~J., {Boulanger}, F., {Stolovy}, S.~R.,
  {Padgett}, D.~L., {Burgdorf}, M.~J., {Fajardo-Acosta}, S.~B., {Glaccum},
  W.~J., {Helou}, G., {Hoard}, D.~W., {Karr}, J., {O'Linger}, J., {Rebull},
  L.~M., {Rho}, J., {Stauffer}, J.~R. \& {Wachter}, S.
\newblock 2004, \ApJS, 154, 281--285.

\bibitem[{Jeong} et~al.(2005)]{jeong2005}
{Jeong}, W.~S., {Mok Lee}, H., {Pak}, S., {Nakagawa}, T., {Minn Kwon}, S.,
  {Pearson}, C.~P. \& {White}, G.~J.
\newblock 2005, \MNRAS, 357, 535--547.

\bibitem[Jewell(2001)]{jewell2001}
Jewell, J.
\newblock 2001, \ApJ, 557, 700.

\bibitem[Joncas et~al.(1992)]{joncas1992}
Joncas, G., Boulanger, F. \& Dewdney, P.~E.
\newblock 1992, \ApJ, 397, 165.

\bibitem[Kiss et~al.(2003)]{kiss2003}
Kiss, C., Abraham, P., Klaas, U., Lemke, D., Heraudeau, P., Burgo, C.~D. \&
  Herbstmeier, U.
\newblock 2003, \AaA, 399, 177.

\bibitem[Lagache et~al.(2000)]{lagache2000a}
Lagache, G., Haffner, L.~M., Reynolds, R.~J. \& Tufte, S.~L.
\newblock 2000, \AaA, 354, 247.

\bibitem[{Li} \& {Draine}(2001)]{li2001}
{Li}, A. \& {Draine}, B.~T.
\newblock 2001, \ApJ, 554, 778--802.

\bibitem[Miville-Desch\^enes \& Lagache(2005)]{miville-deschenes2005a}
Miville-Desch\^enes, M.~A. \& Lagache, G.
\newblock 2005, \ApJS, 157, 302--323.

\bibitem[Miville-Desch\^enes et~al.(2002)]{miville-deschenes2002b}
Miville-Desch\^enes, M.~A., Lagache, G. \& Puget, J.~L.
\newblock 2002, \AaA, 393, 749.

\bibitem[Miville-Desch\^enes et~al.(2003a)]{miville-deschenes2003c}
Miville-Desch\^enes, M.~A., Joncas, G., Falgarone, E. \& Boulanger, F.
\newblock 2003a, \AaA, 411, 109.

\bibitem[Miville-Desch\^enes et~al.(2003b)]{miville-deschenes2003b}
Miville-Desch\^enes, M.~A., Levrier, F. \& Falgarone, E.
\newblock 2003b, \ApJ, 593, 831.

\bibitem[{Padoan} et~al.(2001)]{padoan2001}
{Padoan}, P., {Rosolowsky}, E.~W. \& {Goodman}, A.~A.
\newblock 2001, \ApJ, 547, 862.

\bibitem[Padoan et~al.(2002)]{padoan2002}
Padoan, P., Cambr\'esy, L. \& Langer, W.
\newblock 2002, \ApJ, 580, L57.

\bibitem[Padoan et~al.(2003)]{padoan2003}
Padoan, P., Goodman, A.~A. \& Juvela, M.
\newblock 2003, \ApJ, 588, 881.

\bibitem[{Padoan} et~al.(2006)]{padoan2006}
{Padoan}, P., {Cambr\'esy}, L., {Juvela}, M., {Kritsuk}, A., {Langer}, W.~D. \&
  {Norman}, M.~L.
\newblock 2006, \ApJ, 649, 807--815.

\bibitem[{Rocha} et~al.(2005)]{rocha2005}
{Rocha}, G., {Hobson}, M.~P., {Smith}, S., {Ferreira}, P. \& {Challinor}, A.
\newblock 2005, \MNRAS, 357, 1--11.

\bibitem[Stanimirovic \& Lazarian(2001)]{stanimirovic2001}
Stanimirovic, S. \& Lazarian, A.
\newblock 2001, \ApJ, 551, L53.

\bibitem[Stanimirovic et~al.(1999)]{stanimirovic1999}
Stanimirovic, S., Staveley-Smith, L., Dickey, J.~M., Sault, R.~J. \& Snowden,
  S.~L.
\newblock 1999, \MNRAS, 302, 417.

\bibitem[Starck \& Murtagh(1998)]{starck1998}
Starck, J.~L. \& Murtagh, F.
\newblock 1998, \PASP, 110, 193.

\bibitem[Stepnik et~al.(2003)]{stepnik2003}
Stepnik, B., Abergel, A., Bernard, J.~P., Boulanger, F., Cambr\'esy, L., Giard,
  M., Jones, A.~P., Lagache, G., Lamarre, J.~M., Meny, C., Pajot, F., Peintre,
  F.~L., Ristorcelli, I., Serra, G. \& Torre, J.~P.
\newblock 2003, \AaA, 398, 551.

\bibitem[Stutzki et~al.(1998)]{stutzki1998}
Stutzki, J., Bensch, F., Heithausen, A., Ossenkopf, V. \& Zielinsky, M.
\newblock 1998, \AaA, 336, 697.

\bibitem[{Vio} et~al.(2001)]{vio2001}
{Vio}, R., {Andreani}, P. \& {Wamsteker}, W.
\newblock 2001, \PASP, 113, 1009--1020.

\bibitem[{Vio} et~al.(2002)]{vio2002}
{Vio}, R., {Andreani}, P., {Tenorio}, L. \& {Wamsteker}, W.
\newblock 2002, \PASP, 114, 1281--1289.

\bibitem[Vogelaar \& Wakker(1994)]{vogelaar1994}
Vogelaar, M. G.~R. \& Wakker, B.~P.
\newblock 1994, \AaA, 291, 557.

\bibitem[Wright(1998)]{wright1998}
Wright, E.~L.
\newblock 1998, \ApJ, 496, 1.

\end{thebibliography}
\end{document}